\documentclass[%
 reprint,
 amsmath,amssymb,
 aps,
]{revtex4-1}

\usepackage{graphicx}
\usepackage{dcolumn}
\usepackage{bm}

\begin{document}


\title{Currents and flux-inversion in photokinetic active particles}

\author{Claudio Maggi$^{a}$}
\author{Luca Angelani$^{b}$}
\author{Giacomo Frangipane$^{c}$}
\author{Roberto Di Leonardo$^{a,c}$}

\affiliation{}
\affiliation{\textit{$^{a}$~NANOTEC-CNR, Institute of Nanotechnology, Soft and Living Matter Laboratory - Piazzale A. Moro 2, I-00185, Roma, Italy}}
\affiliation{\textit{$^{b}$~ISC-CNR, Institute for Complex Systems - Piazzale A. Moro 2, I-00185, Roma, Italy}}
\affiliation{\textit{$^{c}$~Dipartimento di Fisica, Universit\`{a} di Roma ``Sapienza'', I-00185, Roma, Italy}}

\date{\today}

\begin{abstract}
Many active particles, both of  biological and synthetic origin, can have a light controllable propulsion speed, a property that in biology is commonly referred to as photokinesis. Here we investigate directed transport of photokinetic particles by traveling light patterns. We find general expressions for the current in the cases where the motility wave, induced by light, shifts very slow or very fast. These asymptotic formulas are independent on the shape of the wave and are valid for a wide class of active particle models. Moreover we derive an exact solution for the one-dimensional ``run and tumble'' model.
Our results could be used to design time-varying illumination patterns for fast and efficient spatial reconfiguration of photokinetic colloids or bacteria.
\end{abstract}

\maketitle


 \section*{Introduction}

Active particles can move with a persistent velocity that has two main components: a drift velocity produced by external forces and a self-propulsion velocity. A considerable amount of research in active matter dynamics has been focused on the effects of non homogeneous force fields~\cite{bechinger2016active} and in particular on the possibility of obtaining directed transport by means of rectification phenomena~\cite{reichhardt2017ratchet}. The emergence of a net particle current, in absence of a net external force, requires the simultaneous breaking of both time reversal and mirror symmetries~\cite{reimann}. In active particles systems this can be simply achieved by an external asymmetric potential that directly breaks mirror symmetry and indirectly breaks time-reversal symmetry by generating, in conjunction with self-propulsion forces, irreversible microscopic trajectories \cite{angelani2011active}.
It has been shown that these rectified currents are potentially exploitable in micro-engineering applications such as micro-cargo delivery~\cite{targeted} and micro-machines actuated by swimming bacteria~\cite{ratch,angelani2009self,sokolov2010swimming,kaiser2014transport,3dmot} or catalytic self-propelled particles~\cite{janus}. 
More recent work has introduced several examples of synthetic self-propelled particles and biological active particles with a self-propulsion speed that is controllable in space and time by shaping light exposure~\cite{busta, palacci2013living, Buttinoni2013, kummel2013circular, palacci2014light}. Swimming \textit{E. coli} cells, expressing a light-driven proton pump, can have a light controllable speed~\cite{busta} which allows to shape their density in response to inhomogeneous illumination~\cite{poon, mona}. However, static light patterns can only prescribe the speed (photokinesis) at each space location thus generating single particle trajectories that obey the same microscopic dynamical laws when viewed in both directions of time. This is not true when an orientational response to light intensity gradients is also present, as in the case of Janus particles for which artificial phototaxis has been recently reported~\cite{bechi}. To see a net current in a system of purely photokinetic active particles we have to break time reversal symmetry at a microscopic level. One possibility is to increase particle density, so that inter-particle interactions become important~\cite{stenhammar2016light}. The other natural route is that of using light patterns evolving with a time asymmetric dynamics.
This idea has been recently investigated in~\cite{hanggi} where it was shown that a shifting (periodic) speed pattern generates a net current in a model system of active Brownian particles. A more interesting finding is that, for slow wave speeds, particles drift in the opposite direction to that of the traveling wave.

Here we show that flow inversion is a general phenomenon for all models of non-interacting active particles having an exponentially correlated propulsion force.
This result holds whenever these particles are subjected to a generic periodic shifting speed pattern of the form $v(x,t)=v(x-c\,t)$ where $x$ is one of the coordinated axes, $c$ is the speed of the wave and $t$ is time.
We find analytical expressions for the current in both the low and high wave speed limits which are independent on: i) the wave profile (sinusoidal, rectangular, ...), ii) the reorientation mechanism (run and tumble, active Brownian, ...), iii) the dimensionality of space (1$d$, 2$d$, 3$d$). 
Finally we obtain an exact solution for 1$d$ ``run and tumble'' particles in a square light intensity wave traveling with an arbitrary speed. We confirm all our theoretical predictions by simulations of various models with different dimensionality and different wave shapes.

It is important to remark that, in photokinetic particles systems, these currents arise in absence of any external force and therefore they are very different from those generated by traveling force fields which have been investigated in driven passive particles~\cite{yellen2009nonlinear,tierno2016transport} and recently also in active particles systems~\cite{sandor2017collective,marconi2017self}.
Our photokinetic particles currents are obtained by controlling only the modulus of velocity in space and time so they are also very different from the chemotactic fluxes that may be generated by a traveling wave of chemoattractant~\cite{goldstein1996traveling}. Indeed in this case concentration gradients determine also the direction of the motion of the particles.  
Our results could be specifically used to design shifting illumination patterns for the optimal transport of photokinetic colloids or bacteria.

\section*{Theory}
We consider the generic equation of motion for the $x$-coordinate of the active particle:
\begin{equation} \label{eqm}
\dot{x} = v(x - c \,t) \, \xi
\end{equation}
\noindent where $v(x)$ has period $\lambda$ and shifts with speed $c$. Here $\xi$ is the ``active noise'' source whose properties will be detailed in the following. By introducing the variable 
${x'=x-c \, t}$ (i.e. by switching to the reference frame of the wave) Eq.~(\ref{eqm}) becomes

\begin{equation} \label{eqm2}
\dot{x'} = v(x') \, \xi - c
\end{equation}

\noindent If we indicate by $P(x',\xi)$ the distribution in phase space, its equation of motion reads

\begin{equation} \label{eqP}
\partial_t P = -\partial_{x'} 
\left[
(v \, \xi - c) \, P
\right]
+ \mathbf{W} \, P
\end{equation}

\noindent where $\mathbf{W}$ is an evolution operator, acting only on $\xi$. We choose $\mathbf{W}$ such that

\begin{eqnarray} 
& & \int d \xi \, \mathbf{W}P=0 \label{oper1} \\
& & \int d \xi \, \xi \, \mathbf{W}P = 
-\alpha \, \langle \xi \rangle_{x'} \, \rho
\label{oper2}
\end{eqnarray}

\noindent where $\rho$ is the marginal probability ${\rho=\int d\xi \, P}$ (i.e. the position probability density), $\langle \xi \rangle_{x'}$ is the average value of the noise at $x'$ and $\alpha$ is the relaxation rate of $\xi$. 
From now on we will indicate the average over $\xi$ with the single brackets, 
${\langle ... \rangle_{x'} = \rho^{-1} \int P(x',\xi) \, ... \, d \xi}$, while the average over both $x$ and $\xi$ will be indicated by the double brackets: ${\langle \langle ... \rangle \rangle = \int \int P(x',\xi) \, ... \, d \xi dx'}$.
Note that, as shown in Ref.~\cite{barriers}, Eq.s~(\ref{oper1}) and (\ref{oper2}) ensure that $\langle\langle \xi \rangle \rangle = 0$ and ${\langle\langle \xi(0) \, \xi(t) \rangle \rangle = \langle  \langle \xi^2 \rangle \rangle \exp(-\alpha \, t)}$ so that the results that follow are generally valid for a broad class of active particles models with exponentially correlated noise. Widely used examples of such models are the ``run and tumble'' (RT) model~\cite{berg2008coli}, the ``active Brownian'' (AB) particle model~\cite{romanczuk2012active} and the ``Gaussian colored noise'' model~\cite{mucna}. 
From now on we consider the stationary state by setting $\partial_t P= 0$. Integrating Eq.~(\ref{eqP})
with respect to $\xi$ we obtain

\begin{equation}\label{eqM0}
-\partial_{x'} J' = -\partial_{x'}
\left[
\rho \, \langle \xi \rangle_{x'} \, v 
-\rho \, c
\right] = 0
\end{equation}

\noindent where we have introduced the density current in the wave reference frame $J'$ which must be a constant. Integrating Eq.~(\ref{eqP}), after multiplication by $\xi$, we have 

\begin{equation}\label{eqM1}
\partial_{x'} 
\left[
\frac{\rho}{v}
\left(
\langle \xi^2 \rangle_{x'} \, v^2 - c^2
\right)
-\frac{c \, J'}{v}
\right]
+\frac{\alpha \, J'}{v}
+\frac{\alpha \, \rho \, c}{v}=0
\end{equation}

\noindent If we now integrate Eq.~({\ref{eqM1}}) in $x'$ over the period $\lambda$ the term in the derivative is zero, since both $\rho$ and $v$ are periodic over $\lambda$, thus we get

\begin{equation}\label{J1g}
J' = -\frac{c}{\lambda} \, \frac{\langle 
\langle 
v^{-1}
\rangle
\rangle}{\overline{v^{-1}}}
\end{equation}

\noindent where we have introduced the ensemble average of $v^{-1}$  

$$
{\langle \langle  v^{-1} \rangle \rangle = \int_0^\lambda dx' \, \frac{\rho}{v}}
$$  

\noindent and the (non-weighted) spatial average 

$$
{\overline{ v^{-1} } = \lambda^{-1} \int_0^\lambda dx'\frac{1}{v}}
$$. 

From Eq.~(\ref{J1g}) we can derive the asymptotic behavior of $J'$ both in the low and high $c$ regimes by properly approximating the probability density $\rho$. For low $c$ we set 

$$
\rho \approx \frac{v^{-1}} {\int_0^\lambda dx' \, v^{-1} }
$$ 

\noindent so that

$$
{\langle \langle v^{-1} \rangle \rangle  
\approx \frac{ \int_0^\lambda dx' \, v^{-2} }
{ \int_0^\lambda dx' \, v^{-1} } }
$$ 

\noindent Thus for low $c$ we have 
${J' \approx -(c/\lambda)
\left( \overline{v^{-2}}/ \, \overline{v^{-1}}^2 \, \right)}$. The current in the laboratory reference frame is obtained as $J = (c/\lambda)+J'$, therefore

\begin{equation}\label{J1sc}
J \approx \frac{c}{\lambda} 
\left(1-
\frac{ 
\overline{v^{-2}}
}
{\overline{v^{-1}}^2}
\right)
\end{equation}

\noindent which is negative since
$\overline{v^{-2}} \geq \overline{v^{-1}}^2$ (Schwartz inequality). This implies that for low $c$ the particles drift in a direction which is opposite with respect to the wave direction.

For high $c$ we can approximate $\rho$ by considering that each active particle takes a long time to reorient and it can ``average'' over many periods of the wave before $\xi$ changes appreciably. In this case the $\xi$ can be considered as ``frozen'' and the density becomes $\rho \propto 1/(c+\xi \, v)$. By normalizing this $\rho$ and expanding to second order in $v$ we find:

$$
\rho \approx 
\lambda^{-1} \left(
1+\frac{\xi ^2 v^2}{c^2}-\frac{\xi ^2 v \overline{v}}{c^2}+\frac{\xi ^2 \overline{v}^2}{c^2}-\frac{\xi ^2
   \overline{v^2}}{c^2}-\frac{\xi  v}{c}+\frac{\xi  \overline{v}}{c}
\right)
$$

\noindent Averaging over $\xi$ we get

\begin{equation} \label{racl}
\rho \approx 
\lambda^{-1}\left[
1+\langle \langle \xi^2 \rangle \rangle \left( 
\frac{ v^2}{c^2}-\frac{ v \overline{v}}{c^2}+\frac{ \overline{v}^2}{c^2}-\frac{
   \overline{v^2}}{c^2}
   \right)
\right]
\end{equation}

\noindent Inserting Eq.~(\ref{racl}) in Eq.~(\ref{J1g}) we obtain for the current
${J'\approx-(c/\lambda)[1-\langle \langle \xi^2 \rangle \rangle (\overline{v^2}-\overline{v}^2)/c^2]}$, therefore in the laboratory reference frame 

\begin{equation}\label{J1lc}
J \approx \frac{\langle \langle \xi^2 \rangle \rangle }{ c \, \lambda} 
\left(
\overline{v^2}-
\overline{v}^2
\right)
\end{equation}

\noindent which is positive, since $\overline{v^2} \geq
\overline{v}^2$, indicating that the particles drift in the direction of the wave at high $c$.
Eq.s~(\ref{J1sc}) and (\ref{J1lc}) are ``universal'', i.e independent on the specific exponentially correlated noise, and prove that there must be a particular $c$ where the flow inversion occurs. Moreover from these two equations is clear that the asymptotic $J$ is independent on the shape of the wave and on the relaxation rate of the noise.

\begin{figure}
 \centering
 \includegraphics[width=.99\columnwidth]{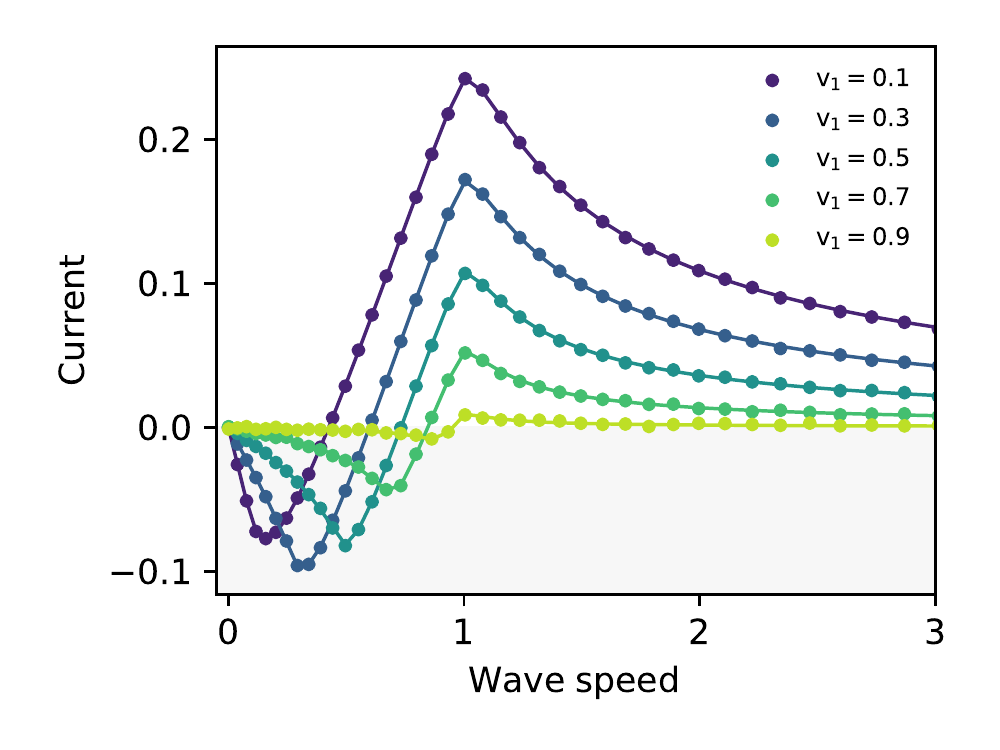}
  \caption{Current as function of the wave speed in simulations (circles) and theory (full lines) for the RT1$d$ model. Different colors represent $J$ for various values of $v_1$ (see legend). Here we fix $\alpha=1$. 
  The shaded area highlights the negative current region.}
  \label{f1}
\end{figure}

\begin{figure}
 \centering
 \includegraphics[width=.99\columnwidth]{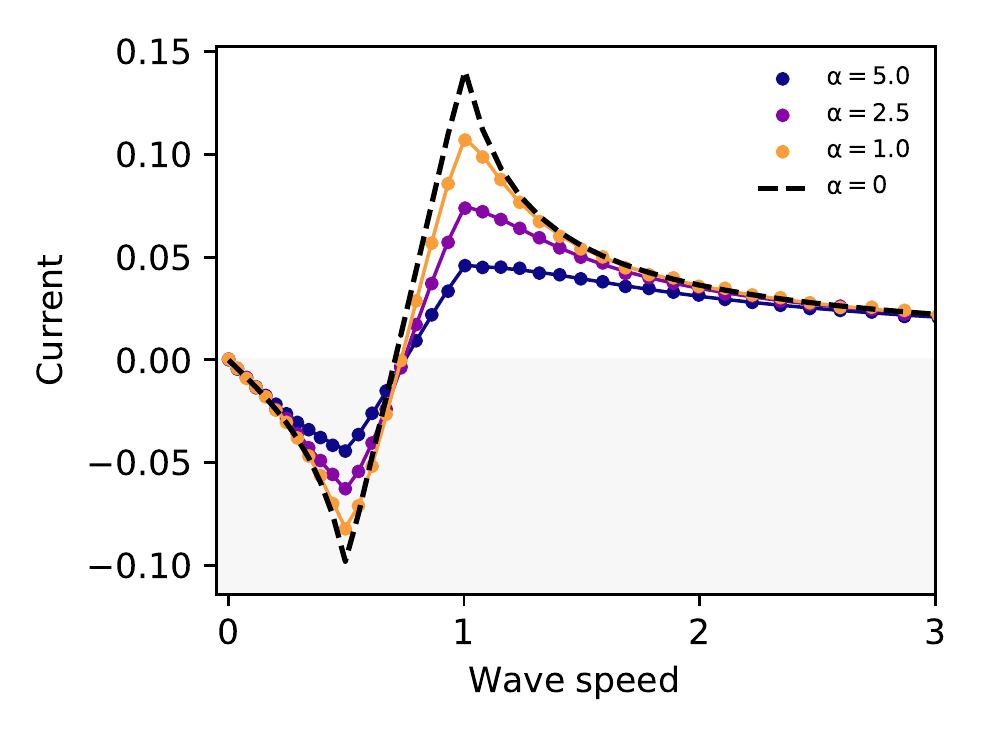}
\caption{Current as function of the wave speed in simulations (circles) and theory (full lines) for the RT1$d$ model. Different colors represent $J$ for various values of $\alpha$ (see legend). 
The dashed line is the theoretical curve for smooth-swimmers ($\alpha=0$). Here we fix $v_1=1/2$. 
The shaded area highlights the negative current region.}
\label{f2}
\end{figure}

We now derive the full analytic solution for $J$ for all values of $c$ for the RT model in 1d, for  a square wave. In this case Eq.~(\ref{eqM1}) becomes

\begin{equation} \label{Jstat}
J' \left( \frac{\alpha}{v} -  c\ \partial_{x'} \frac1v \right) = 
- \partial_{x'} \left( \frac{v^2-c^2}{v} \, \rho  \right) - \frac{\alpha c}{v} \rho
\end{equation}

\noindent since $\langle \xi^2 \rangle_{x'} = 1$. Specializing the calculation to the case of a square wave we consider
$ v(x)=v_1$ if $0< x < \lambda/2$ and $v(x)=v_2$  if $\lambda/2 < x < \lambda$.
By using the same procedure adopted in Ref.~\cite{angelani2011active},
we solve  Eq.~(\ref{Jstat}) for the probabilities $P_n$ ($n=1,2$) in the two regions and imposing normalization and boundary discontinuous conditions
at $0$ and $\lambda/2$, we finally obtain an expression for $J$.
For $c<v_1$ and $c>v_2$ the expressions for the current $J$ in the laboratory frame is

\begin{eqnarray} \label{J1}
&&J = \frac{c}{\lambda} \left[
\frac{\sinh (A_1+A_2)} {K \  \sinh A_1 \sinh A_2 - \sinh (A_1+A_2)}
+1 \right]\\\nonumber
&&A_1 =\  \frac{\alpha c}{4} \frac{\lambda}{v_1^2-c^2}\;,\;\;\;\; A_2 =  \frac{\alpha c}{4} \frac{\lambda}{v_2^2-c^2}\\\nonumber
&&K =\frac{ 2 (v_2-v_1)^2}{\alpha \lambda c}
\end{eqnarray}

\noindent 
For $v_1<c<v_2$ we have to consider that, in the traveling frame,  right-oriented particles accumulate at $x'=0$. To understand this we note that the 
the velocity field of right-oriented particles is $v(x')-c$ that changes sign at $x'=0$ and $x'=\lambda/2$. Such velocity field points towards ${x'=0}$  both from the left and from the right (see Supplementary Figure), while it points away from $x'=\lambda/2$.
This builds up a $\delta$-function contribution to the probability density at $x'=0$.
Considering this, after some algebra, we obtain

\begin{eqnarray}\label{J2}
&&J = \frac{c}{\lambda} \left[
\frac{1} {
Q_2 (1-e^{-2 A_2}) 
- Q_1 (1- e^{2 A_1}) 
- 1}
+1 \right]\\\nonumber
&&Q_1= \frac{(c-v_1)^2}{\alpha \lambda c}\;,\;\;\;Q_2= \frac{(v_2-c)^2}{\alpha \lambda c}
\end{eqnarray}

\noindent In the case of ``smooth swimmers'', i.e. non-tumbling particles, we can derive a simpler expression of the current $J$
by taking the limit $\alpha \to 0$ in Eq.s~(\ref{J1}) and (\ref{J2}). We have, for $c<v_1$ and $c>v_2$

\begin{equation} \label{smo1}
J_{ss} = \frac{c}{\lambda} \ 
\frac{(v_2-v_1)^2}{4c^2-(v_1+v_2)^2}
\end{equation}

\noindent and for $v_1<c<v_2$

\begin{equation}\label{smo2}
J_{ss} = \frac{c}{\lambda} \ 
\frac{ c- v_1 v_2/c}
{2c+v_1+v_2}
\end{equation}

\section*{Simulations}
\begin{figure}[h]
 \centering
 \includegraphics[width=.99\columnwidth]{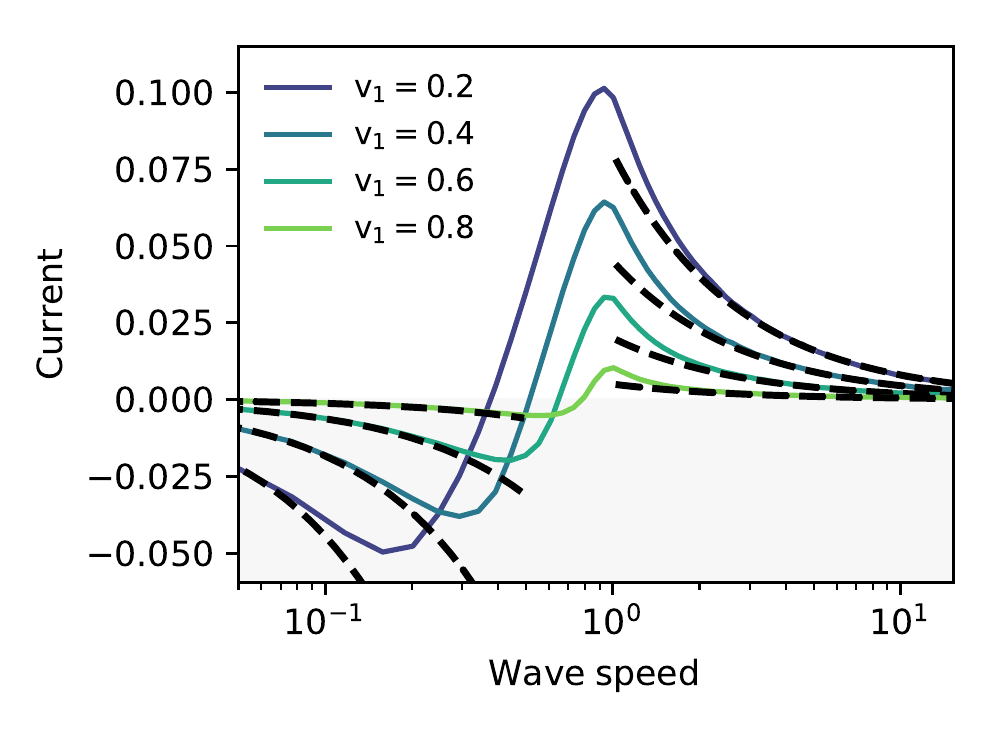}
\caption{Current as function of the wave speed in $2d$ simulations of the RT model (full lines), different colors represent $J$ for various values of $v_1$ (see legend). Dashed lines are the theoretical prediction for asymptotes. Here we fix $\alpha=1$. 
The shaded area highlights the negative current region.}
\label{f3}
\end{figure}

\begin{figure}[h]
 \centering
 \includegraphics[width=.99\columnwidth]{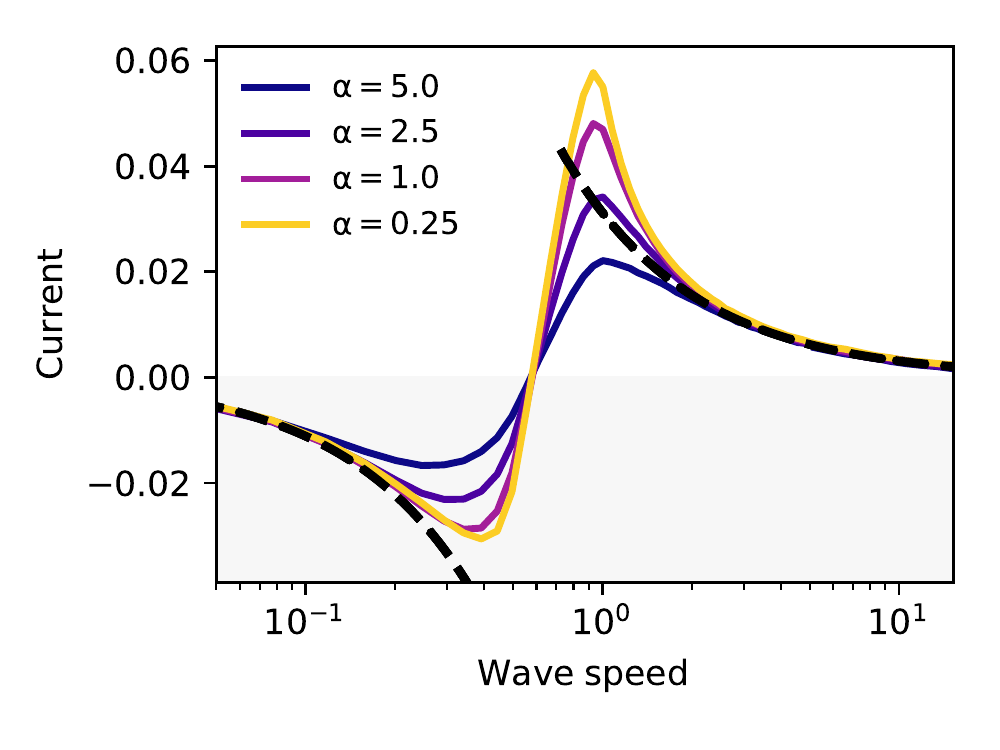}
\caption{Current as function of the wave speed in $2d$ simulations of the RT model (full lines), different colors represent $J$ for various values of $\alpha$ (see legend). Dashed lines are the theoretical prediction for asymptotes. Here we fix $v_1=1/2$. 
The shaded area highlights the negative current region.}
\label{f4}
\end{figure}

To verify the results of the previous section 
we perform computer simulations of various models in different dimensions and with different wave shapes. In all our numerical calculations we integrate directly the stochastic equation of motion (\ref{eqm2}), for non-interacting active particles. We use a GPU-based simulation that allows to evolve tenths of thousands of particles, for millions of time-steps, in few minutes.

We start by considering the RT model in 1d to check the analytic formulas (\ref{J1}) and (\ref{J2}) for the square wave. In this case $\xi$ is telegraphic noise that switches between $-1$ and $+1$ at a rate $\alpha$. In these simulations we fix the period of the wave $\lambda=1$ and square wave maximum speed $v_2=1$. In Fig.~\ref{f1} we show the theory and the simulation results for several different values of the low speed value $v_1$. As it can be seen the agreement between theory and simulation is perfect validating our analytic result. Moreover it is clear that as $v_1$ decreases $J$ increases in the $c>v_2$ region while it increases and then decreases in modulus at low $c$. In Fig.~\ref{f2} we show $J$ as a function of $c$ at fixed $v_1=1/2$ upon varying $\alpha$. It is evident that $|J|$ increases both in the low and high  $c$ regimes up to a limiting value represented by the $\alpha=0$ case.

To check the validity of the asymptotic formulas (\ref{J1sc}) and (\ref{J1lc}) we run simulations of the RT model in 2d subjected to a square wave speed pattern. In this case $\xi$ in Eq.~(\ref{eqm2}) becomes $\xi=\cos(\theta)$, where the orientation angle $\theta$ changes abruptly during ``tumble'' events (happening at rate $\alpha$) taking a random value between $0$ and $2 \pi$ (note that here $\langle \langle \xi^2 \rangle \rangle = 1/2 \,$). In Fig.~\ref{f3} we show the asymptotic theory and the simulation results varying $v_1$. We see that the effect of varying $v_1$ is qualitatively very similar to the 1d case and that the asymptotic expressions (\ref{J1sc}) and (\ref{J1lc}) capture $J$ at low and high $c$. In the case where $v_1$ and $v_2$ are fixed and $\alpha$ varies (Fig.~\ref{f4}) we find again a strong analogy with the 1d case, i.e. $|J|$ increases as $\alpha$ decreases at all values of $c$, moreover Eq.s~(\ref{J1sc}) and (\ref{J1lc}) result to be correct asymptotically.

\begin{figure}
 \centering
 \includegraphics[width=.99\columnwidth]{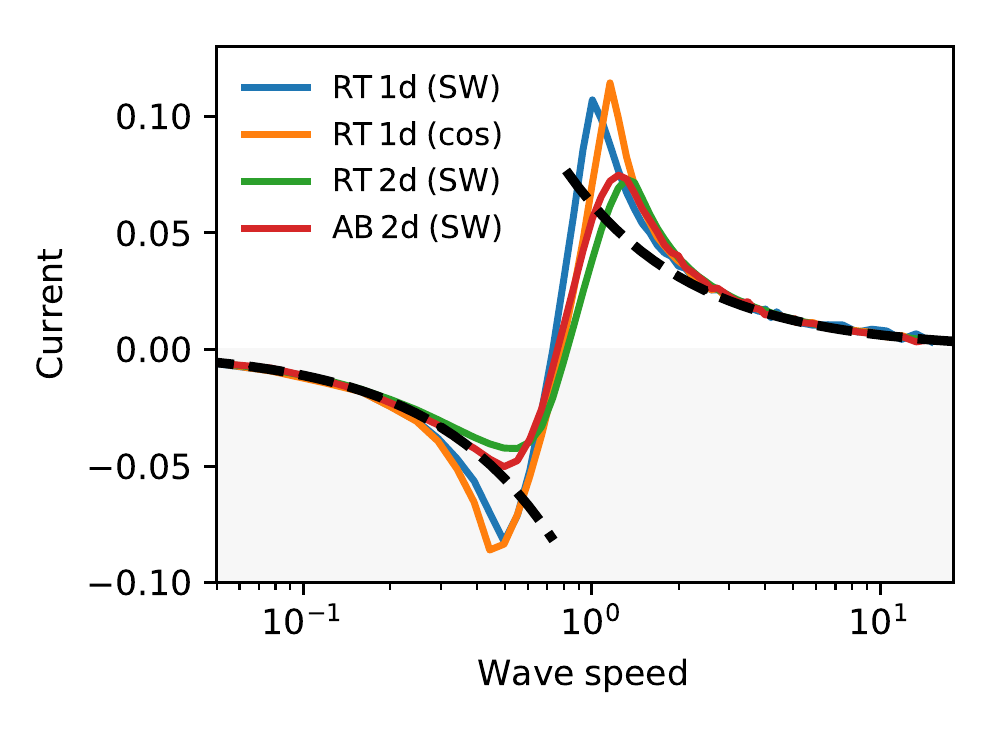}
\caption{The colored curves represent the current in simulations as function of the wave speed in different models, wave shapes and dimensionality (see legend). Here we fix $\alpha = 1$ for all models and we choose $v(x)$ 
so that the variance of the speed and the ratio $\overline{v^{-2}}/\overline{v^{-1}}^2$ is the same in all cases. 
The dashed lines are the asymptotes from theory.}
\label{f5}
\end{figure}

For testing the independence of the results (\ref{J1sc}) and (\ref{J1lc}) on the particular form of $v(x)$ we also run simulation of the 1d RT model in the case where $v$ is a cosine with maximum and minimum speed values $v_b$ and $v_a$, i.e. ${v(x) = (v_b-v_a) [\cos(2 \pi x/\lambda)+1]/2+v_a}$. We choose $v_a$ and $v_b$ such that $\left(\overline{v^2}-\overline{v}^2\right)$ and $\left(\overline{v^{-2}}/\overline{v^{-1}}^2\right)$ computed for he cosine are the same of those of the square wave used above with $v_1=1/2$ and $v_2=1$.
In this way, according to (\ref{J1sc}) and (\ref{J1lc}), the asymptotes of $J$ for the two cases should coincide. In Fig.~\ref{f5} we see that the current for the two $v(x)$ collapse on the asymptotic curves both in the low and high $c$ regions. This ``master plot'' can be enriched by adding the numerical results of the 2d RT model in the square wave case. If we choose $v_1=\sqrt{2}/2$ and $v_2=\sqrt{2}$ then the asymptotic forms of the 1d and 2d cases should be identical absorbing the factor $\langle \langle \xi^2 \rangle \rangle = 1/2$, as verified in Fig.~\ref{f5}. 

Finally we also check the validity of these asymptotes for AB particles. In this model $\xi = \cos(\theta)$, as in the 2d RT case,
but $\theta$ evolves more gradually because rotational Brownian motion. For AB particles 
$\dot{ \theta } = \alpha^{-1} \, \eta$, where $\eta$ is standard (delta-correlated) white noise: $\langle \eta(0)\eta(t) \rangle = 2 \delta(t)$. We use a square wave for $v(x)$ with $v_1=\sqrt{2}/2$ and $v_2=\sqrt{2}$ as in the RT case above. Again in Fig.~\ref{f5} we show that the numerical asymptotes fall on top of the other curves.

\section*{Conclusions}
We have derived the asymptotic expressions for the current in active particles models with exponentially correlated self-propulsion, when these are subjected to traveling periodic speed landscapes.
Our equations show that, when the traveling pattern moves slowly with respect to the particle minimum speed, the flow occurs oppositely to the wave shifting direction. Differently, when the pattern shifts faster than the particle maximum speed, the current goes in the same direction of the traveling wave. This demonstrates that flow inversion is a general phenomenon occurring in a wide class of active particles models.
Moreover our asymptotic formulas are independent on the shape of the wave and on the propulsion relaxation in the specific model as confirmed by computer simulations. Finally we have derived an exact solution for the ``run and tumble'' model in one dimension in the case where the speed profile is a square wave. This model gives valuable information about how the current increases upon increasing the relaxation time of the propulsion and allows to know precisely which pattern speed maximize the positive or negative current. 
It could be also interesting to perform simulations of interacting particles~\cite{angelani2009self} subjected to shifting speed patterns. To describe these systems theoretically one could start by using schematic lattice models with excluded volume interactions~\cite{chatterjee2014interacting}.

Our results provide a strategy to design traveling light patterns with shape and speed that are optimized for the transport of photokinetic bacteria or self-propelled colloids. From a practical point of view these currents could be used for the delivery of colloidal cargoes in  target regions that are reconfigurable and without the need of permanent micro-fabricated structures. 

\section*{Acknowledgement}
We acknowledge NVIDIA for hardware donation.

\bibliography{references}

\begin{thebibliography}{31}%
\makeatletter
\providecommand \@ifxundefined [1]{%
 \@ifx{#1\undefined}
}%
\providecommand \@ifnum [1]{%
 \ifnum #1\expandafter \@firstoftwo
 \else \expandafter \@secondoftwo
 \fi
}%
\providecommand \@ifx [1]{%
 \ifx #1\expandafter \@firstoftwo
 \else \expandafter \@secondoftwo
 \fi
}%
\providecommand \natexlab [1]{#1}%
\providecommand \enquote  [1]{``#1''}%
\providecommand \bibnamefont  [1]{#1}%
\providecommand \bibfnamefont [1]{#1}%
\providecommand \citenamefont [1]{#1}%
\providecommand \href@noop [0]{\@secondoftwo}%
\providecommand \href [0]{\begingroup \@sanitize@url \@href}%
\providecommand \@href[1]{\@@startlink{#1}\@@href}%
\providecommand \@@href[1]{\endgroup#1\@@endlink}%
\providecommand \@sanitize@url [0]{\catcode `\\12\catcode `\$12\catcode
  `\&12\catcode `\#12\catcode `\^12\catcode `\_12\catcode `\%12\relax}%
\providecommand \@@startlink[1]{}%
\providecommand \@@endlink[0]{}%
\providecommand \url  [0]{\begingroup\@sanitize@url \@url }%
\providecommand \@url [1]{\endgroup\@href {#1}{\urlprefix }}%
\providecommand \urlprefix  [0]{URL }%
\providecommand \Eprint [0]{\href }%
\providecommand \doibase [0]{http://dx.doi.org/}%
\providecommand \selectlanguage [0]{\@gobble}%
\providecommand \bibinfo  [0]{\@secondoftwo}%
\providecommand \bibfield  [0]{\@secondoftwo}%
\providecommand \translation [1]{[#1]}%
\providecommand \BibitemOpen [0]{}%
\providecommand \bibitemStop [0]{}%
\providecommand \bibitemNoStop [0]{.\EOS\space}%
\providecommand \EOS [0]{\spacefactor3000\relax}%
\providecommand \BibitemShut  [1]{\csname bibitem#1\endcsname}%
\let\auto@bib@innerbib\@empty
\bibitem [{\citenamefont {Bechinger}\ \emph {et~al.}(2016)\citenamefont
  {Bechinger}, \citenamefont {Di~Leonardo}, \citenamefont {L{\"o}wen},
  \citenamefont {Reichhardt}, \citenamefont {Volpe},\ and\ \citenamefont
  {Volpe}}]{bechinger2016active}%
  \BibitemOpen
  \bibfield  {author} {\bibinfo {author} {\bibfnamefont {C.}~\bibnamefont
  {Bechinger}}, \bibinfo {author} {\bibfnamefont {R.}~\bibnamefont
  {Di~Leonardo}}, \bibinfo {author} {\bibfnamefont {H.}~\bibnamefont
  {L{\"o}wen}}, \bibinfo {author} {\bibfnamefont {C.}~\bibnamefont
  {Reichhardt}}, \bibinfo {author} {\bibfnamefont {G.}~\bibnamefont {Volpe}}, \
  and\ \bibinfo {author} {\bibfnamefont {G.}~\bibnamefont {Volpe}},\
  }\href@noop {} {\bibfield  {journal} {\bibinfo  {journal} {Reviews of Modern
  Physics}\ }\textbf {\bibinfo {volume} {88}},\ \bibinfo {pages} {045006}
  (\bibinfo {year} {2016})}\BibitemShut {NoStop}%
\bibitem [{\citenamefont {Reichhardt}\ and\ \citenamefont
  {Reichhardt}(2017)}]{reichhardt2017ratchet}%
  \BibitemOpen
  \bibfield  {author} {\bibinfo {author} {\bibfnamefont {C.~O.}\ \bibnamefont
  {Reichhardt}}\ and\ \bibinfo {author} {\bibfnamefont {C.}~\bibnamefont
  {Reichhardt}},\ }\href@noop {} {\bibfield  {journal} {\bibinfo  {journal}
  {Annual Review of Condensed Matter Physics}\ }\textbf {\bibinfo {volume}
  {8}},\ \bibinfo {pages} {51} (\bibinfo {year} {2017})}\BibitemShut {NoStop}%
\bibitem [{\citenamefont {Reimann}(2002)}]{reimann}%
  \BibitemOpen
  \bibfield  {author} {\bibinfo {author} {\bibfnamefont {P.}~\bibnamefont
  {Reimann}},\ }\href@noop {} {\bibfield  {journal} {\bibinfo  {journal}
  {Physics reports}\ }\textbf {\bibinfo {volume} {361}},\ \bibinfo {pages} {57}
  (\bibinfo {year} {2002})}\BibitemShut {NoStop}%
\bibitem [{\citenamefont {Angelani}\ \emph {et~al.}(2011)\citenamefont
  {Angelani}, \citenamefont {Costanzo},\ and\ \citenamefont
  {Di~Leonardo}}]{angelani2011active}%
  \BibitemOpen
  \bibfield  {author} {\bibinfo {author} {\bibfnamefont {L.}~\bibnamefont
  {Angelani}}, \bibinfo {author} {\bibfnamefont {A.}~\bibnamefont {Costanzo}},
  \ and\ \bibinfo {author} {\bibfnamefont {R.}~\bibnamefont {Di~Leonardo}},\
  }\href@noop {} {\bibfield  {journal} {\bibinfo  {journal} {EPL (Europhysics
  Letters)}\ }\textbf {\bibinfo {volume} {96}},\ \bibinfo {pages} {68002}
  (\bibinfo {year} {2011})}\BibitemShut {NoStop}%
\bibitem [{\citenamefont {Koumakis}\ \emph {et~al.}(2013)\citenamefont
  {Koumakis}, \citenamefont {Lepore}, \citenamefont {Maggi},\ and\
  \citenamefont {Di~Leonardo}}]{targeted}%
  \BibitemOpen
  \bibfield  {author} {\bibinfo {author} {\bibfnamefont {N.}~\bibnamefont
  {Koumakis}}, \bibinfo {author} {\bibfnamefont {A.}~\bibnamefont {Lepore}},
  \bibinfo {author} {\bibfnamefont {C.}~\bibnamefont {Maggi}}, \ and\ \bibinfo
  {author} {\bibfnamefont {R.}~\bibnamefont {Di~Leonardo}},\ }\href@noop {}
  {\bibfield  {journal} {\bibinfo  {journal} {Nature communications}\ }\textbf
  {\bibinfo {volume} {4}},\ \bibinfo {pages} {2588} (\bibinfo {year}
  {2013})}\BibitemShut {NoStop}%
\bibitem [{\citenamefont {Di~Leonardo}\ \emph {et~al.}(2010)\citenamefont
  {Di~Leonardo}, \citenamefont {Angelani}, \citenamefont {Dell'Arciprete},
  \citenamefont {Ruocco}, \citenamefont {Iebba}, \citenamefont {Schippa},
  \citenamefont {Conte}, \citenamefont {Mecarini}, \citenamefont {De~Angelis},\
  and\ \citenamefont {Di~Fabrizio}}]{ratch}%
  \BibitemOpen
  \bibfield  {author} {\bibinfo {author} {\bibfnamefont {R.}~\bibnamefont
  {Di~Leonardo}}, \bibinfo {author} {\bibfnamefont {L.}~\bibnamefont
  {Angelani}}, \bibinfo {author} {\bibfnamefont {D.}~\bibnamefont
  {Dell'Arciprete}}, \bibinfo {author} {\bibfnamefont {G.}~\bibnamefont
  {Ruocco}}, \bibinfo {author} {\bibfnamefont {V.}~\bibnamefont {Iebba}},
  \bibinfo {author} {\bibfnamefont {S.}~\bibnamefont {Schippa}}, \bibinfo
  {author} {\bibfnamefont {M.}~\bibnamefont {Conte}}, \bibinfo {author}
  {\bibfnamefont {F.}~\bibnamefont {Mecarini}}, \bibinfo {author}
  {\bibfnamefont {F.}~\bibnamefont {De~Angelis}}, \ and\ \bibinfo {author}
  {\bibfnamefont {E.}~\bibnamefont {Di~Fabrizio}},\ }\href@noop {} {\bibfield
  {journal} {\bibinfo  {journal} {Proceedings of the National Academy of
  Sciences}\ }\textbf {\bibinfo {volume} {107}},\ \bibinfo {pages} {9541}
  (\bibinfo {year} {2010})}\BibitemShut {NoStop}%
\bibitem [{\citenamefont {Angelani}\ \emph {et~al.}(2009)\citenamefont
  {Angelani}, \citenamefont {Di~Leonardo},\ and\ \citenamefont
  {Ruocco}}]{angelani2009self}%
  \BibitemOpen
  \bibfield  {author} {\bibinfo {author} {\bibfnamefont {L.}~\bibnamefont
  {Angelani}}, \bibinfo {author} {\bibfnamefont {R.}~\bibnamefont
  {Di~Leonardo}}, \ and\ \bibinfo {author} {\bibfnamefont {G.}~\bibnamefont
  {Ruocco}},\ }\href@noop {} {\bibfield  {journal} {\bibinfo  {journal}
  {Physical review letters}\ }\textbf {\bibinfo {volume} {102}},\ \bibinfo
  {pages} {048104} (\bibinfo {year} {2009})}\BibitemShut {NoStop}%
\bibitem [{\citenamefont {Sokolov}\ \emph {et~al.}(2010)\citenamefont
  {Sokolov}, \citenamefont {Apodaca}, \citenamefont {Grzybowski},\ and\
  \citenamefont {Aranson}}]{sokolov2010swimming}%
  \BibitemOpen
  \bibfield  {author} {\bibinfo {author} {\bibfnamefont {A.}~\bibnamefont
  {Sokolov}}, \bibinfo {author} {\bibfnamefont {M.~M.}\ \bibnamefont
  {Apodaca}}, \bibinfo {author} {\bibfnamefont {B.~A.}\ \bibnamefont
  {Grzybowski}}, \ and\ \bibinfo {author} {\bibfnamefont {I.~S.}\ \bibnamefont
  {Aranson}},\ }\href@noop {} {\bibfield  {journal} {\bibinfo  {journal}
  {Proceedings of the National Academy of Sciences}\ }\textbf {\bibinfo
  {volume} {107}},\ \bibinfo {pages} {969} (\bibinfo {year}
  {2010})}\BibitemShut {NoStop}%
\bibitem [{\citenamefont {Kaiser}\ \emph {et~al.}(2014)\citenamefont {Kaiser},
  \citenamefont {Peshkov}, \citenamefont {Sokolov}, \citenamefont {ten Hagen},
  \citenamefont {L{\"o}wen},\ and\ \citenamefont
  {Aranson}}]{kaiser2014transport}%
  \BibitemOpen
  \bibfield  {author} {\bibinfo {author} {\bibfnamefont {A.}~\bibnamefont
  {Kaiser}}, \bibinfo {author} {\bibfnamefont {A.}~\bibnamefont {Peshkov}},
  \bibinfo {author} {\bibfnamefont {A.}~\bibnamefont {Sokolov}}, \bibinfo
  {author} {\bibfnamefont {B.}~\bibnamefont {ten Hagen}}, \bibinfo {author}
  {\bibfnamefont {H.}~\bibnamefont {L{\"o}wen}}, \ and\ \bibinfo {author}
  {\bibfnamefont {I.~S.}\ \bibnamefont {Aranson}},\ }\href@noop {} {\bibfield
  {journal} {\bibinfo  {journal} {Physical review letters}\ }\textbf {\bibinfo
  {volume} {112}},\ \bibinfo {pages} {158101} (\bibinfo {year}
  {2014})}\BibitemShut {NoStop}%
\bibitem [{\citenamefont {Vizsnyiczai}\ \emph {et~al.}(2017)\citenamefont
  {Vizsnyiczai}, \citenamefont {Frangipane}, \citenamefont {Maggi},
  \citenamefont {Saglimbeni}, \citenamefont {Bianchi},\ and\ \citenamefont
  {Di~Leonardo}}]{3dmot}%
  \BibitemOpen
  \bibfield  {author} {\bibinfo {author} {\bibfnamefont {G.}~\bibnamefont
  {Vizsnyiczai}}, \bibinfo {author} {\bibfnamefont {G.}~\bibnamefont
  {Frangipane}}, \bibinfo {author} {\bibfnamefont {C.}~\bibnamefont {Maggi}},
  \bibinfo {author} {\bibfnamefont {F.}~\bibnamefont {Saglimbeni}}, \bibinfo
  {author} {\bibfnamefont {S.}~\bibnamefont {Bianchi}}, \ and\ \bibinfo
  {author} {\bibfnamefont {R.}~\bibnamefont {Di~Leonardo}},\ }\href@noop {}
  {\bibfield  {journal} {\bibinfo  {journal} {Nature communications}\ }\textbf
  {\bibinfo {volume} {8}},\ \bibinfo {pages} {15974} (\bibinfo {year}
  {2017})}\BibitemShut {NoStop}%
\bibitem [{\citenamefont {Maggi}\ \emph {et~al.}(2016)\citenamefont {Maggi},
  \citenamefont {Simmchen}, \citenamefont {Saglimbeni}, \citenamefont {Katuri},
  \citenamefont {Dipalo}, \citenamefont {De~Angelis}, \citenamefont {Sanchez},\
  and\ \citenamefont {Di~Leonardo}}]{janus}%
  \BibitemOpen
  \bibfield  {author} {\bibinfo {author} {\bibfnamefont {C.}~\bibnamefont
  {Maggi}}, \bibinfo {author} {\bibfnamefont {J.}~\bibnamefont {Simmchen}},
  \bibinfo {author} {\bibfnamefont {F.}~\bibnamefont {Saglimbeni}}, \bibinfo
  {author} {\bibfnamefont {J.}~\bibnamefont {Katuri}}, \bibinfo {author}
  {\bibfnamefont {M.}~\bibnamefont {Dipalo}}, \bibinfo {author} {\bibfnamefont
  {F.}~\bibnamefont {De~Angelis}}, \bibinfo {author} {\bibfnamefont
  {S.}~\bibnamefont {Sanchez}}, \ and\ \bibinfo {author} {\bibfnamefont
  {R.}~\bibnamefont {Di~Leonardo}},\ }\href@noop {} {\bibfield  {journal}
  {\bibinfo  {journal} {Small}\ }\textbf {\bibinfo {volume} {12}},\ \bibinfo
  {pages} {446} (\bibinfo {year} {2016})}\BibitemShut {NoStop}%
\bibitem [{\citenamefont {Walter}\ \emph {et~al.}(2007)\citenamefont {Walter},
  \citenamefont {Greenfield}, \citenamefont {Bustamante},\ and\ \citenamefont
  {Liphardt}}]{busta}%
  \BibitemOpen
  \bibfield  {author} {\bibinfo {author} {\bibfnamefont {J.~M.}\ \bibnamefont
  {Walter}}, \bibinfo {author} {\bibfnamefont {D.}~\bibnamefont {Greenfield}},
  \bibinfo {author} {\bibfnamefont {C.}~\bibnamefont {Bustamante}}, \ and\
  \bibinfo {author} {\bibfnamefont {J.}~\bibnamefont {Liphardt}},\ }\href@noop
  {} {\bibfield  {journal} {\bibinfo  {journal} {Proceedings of the National
  Academy of Sciences}\ }\textbf {\bibinfo {volume} {104}},\ \bibinfo {pages}
  {2408} (\bibinfo {year} {2007})}\BibitemShut {NoStop}%
\bibitem [{\citenamefont {Palacci}\ \emph {et~al.}(2013)\citenamefont
  {Palacci}, \citenamefont {Sacanna}, \citenamefont {Steinberg}, \citenamefont
  {Pine},\ and\ \citenamefont {Chaikin}}]{palacci2013living}%
  \BibitemOpen
  \bibfield  {author} {\bibinfo {author} {\bibfnamefont {J.}~\bibnamefont
  {Palacci}}, \bibinfo {author} {\bibfnamefont {S.}~\bibnamefont {Sacanna}},
  \bibinfo {author} {\bibfnamefont {A.~P.}\ \bibnamefont {Steinberg}}, \bibinfo
  {author} {\bibfnamefont {D.~J.}\ \bibnamefont {Pine}}, \ and\ \bibinfo
  {author} {\bibfnamefont {P.~M.}\ \bibnamefont {Chaikin}},\ }\href@noop {}
  {\bibfield  {journal} {\bibinfo  {journal} {Science}\ ,\ \bibinfo {pages}
  {1230020}} (\bibinfo {year} {2013})}\BibitemShut {NoStop}%
\bibitem [{\citenamefont {Buttinoni}\ \emph {et~al.}(2013)\citenamefont
  {Buttinoni}, \citenamefont {Bialk{\'e}}, \citenamefont {K{\"u}mmel},
  \citenamefont {L{\"o}wen}, \citenamefont {Bechinger},\ and\ \citenamefont
  {Speck}}]{Buttinoni2013}%
  \BibitemOpen
  \bibfield  {author} {\bibinfo {author} {\bibfnamefont {I.}~\bibnamefont
  {Buttinoni}}, \bibinfo {author} {\bibfnamefont {J.}~\bibnamefont
  {Bialk{\'e}}}, \bibinfo {author} {\bibfnamefont {F.}~\bibnamefont
  {K{\"u}mmel}}, \bibinfo {author} {\bibfnamefont {H.}~\bibnamefont
  {L{\"o}wen}}, \bibinfo {author} {\bibfnamefont {C.}~\bibnamefont
  {Bechinger}}, \ and\ \bibinfo {author} {\bibfnamefont {T.}~\bibnamefont
  {Speck}},\ }\href@noop {} {\bibfield  {journal} {\bibinfo  {journal}
  {Physical review letters}\ }\textbf {\bibinfo {volume} {110}},\ \bibinfo
  {pages} {238301} (\bibinfo {year} {2013})}\BibitemShut {NoStop}%
\bibitem [{\citenamefont {K{\"u}mmel}\ \emph {et~al.}(2013)\citenamefont
  {K{\"u}mmel}, \citenamefont {ten Hagen}, \citenamefont {Wittkowski},
  \citenamefont {Buttinoni}, \citenamefont {Eichhorn}, \citenamefont {Volpe},
  \citenamefont {L{\"o}wen},\ and\ \citenamefont
  {Bechinger}}]{kummel2013circular}%
  \BibitemOpen
  \bibfield  {author} {\bibinfo {author} {\bibfnamefont {F.}~\bibnamefont
  {K{\"u}mmel}}, \bibinfo {author} {\bibfnamefont {B.}~\bibnamefont {ten
  Hagen}}, \bibinfo {author} {\bibfnamefont {R.}~\bibnamefont {Wittkowski}},
  \bibinfo {author} {\bibfnamefont {I.}~\bibnamefont {Buttinoni}}, \bibinfo
  {author} {\bibfnamefont {R.}~\bibnamefont {Eichhorn}}, \bibinfo {author}
  {\bibfnamefont {G.}~\bibnamefont {Volpe}}, \bibinfo {author} {\bibfnamefont
  {H.}~\bibnamefont {L{\"o}wen}}, \ and\ \bibinfo {author} {\bibfnamefont
  {C.}~\bibnamefont {Bechinger}},\ }\href@noop {} {\bibfield  {journal}
  {\bibinfo  {journal} {Physical review letters}\ }\textbf {\bibinfo {volume}
  {110}},\ \bibinfo {pages} {198302} (\bibinfo {year} {2013})}\BibitemShut
  {NoStop}%
\bibitem [{\citenamefont {Palacci}\ \emph {et~al.}(2014)\citenamefont
  {Palacci}, \citenamefont {Sacanna}, \citenamefont {Kim}, \citenamefont {Yi},
  \citenamefont {Pine},\ and\ \citenamefont {Chaikin}}]{palacci2014light}%
  \BibitemOpen
  \bibfield  {author} {\bibinfo {author} {\bibfnamefont {J.}~\bibnamefont
  {Palacci}}, \bibinfo {author} {\bibfnamefont {S.}~\bibnamefont {Sacanna}},
  \bibinfo {author} {\bibfnamefont {S.-H.}\ \bibnamefont {Kim}}, \bibinfo
  {author} {\bibfnamefont {G.-R.}\ \bibnamefont {Yi}}, \bibinfo {author}
  {\bibfnamefont {D.}~\bibnamefont {Pine}}, \ and\ \bibinfo {author}
  {\bibfnamefont {P.}~\bibnamefont {Chaikin}},\ }\href@noop {} {\bibfield
  {journal} {\bibinfo  {journal} {Phil. Trans. R. Soc. A}\ }\textbf {\bibinfo
  {volume} {372}},\ \bibinfo {pages} {20130372} (\bibinfo {year}
  {2014})}\BibitemShut {NoStop}%
\bibitem [{\citenamefont {Arlt}\ \emph {et~al.}(2018)\citenamefont {Arlt},
  \citenamefont {Martinez}, \citenamefont {Dawson}, \citenamefont {Pilizota},\
  and\ \citenamefont {Poon}}]{poon}%
  \BibitemOpen
  \bibfield  {author} {\bibinfo {author} {\bibfnamefont {J.}~\bibnamefont
  {Arlt}}, \bibinfo {author} {\bibfnamefont {V.~A.}\ \bibnamefont {Martinez}},
  \bibinfo {author} {\bibfnamefont {A.}~\bibnamefont {Dawson}}, \bibinfo
  {author} {\bibfnamefont {T.}~\bibnamefont {Pilizota}}, \ and\ \bibinfo
  {author} {\bibfnamefont {W.~C.}\ \bibnamefont {Poon}},\ }\href@noop {}
  {\bibfield  {journal} {\bibinfo  {journal} {Nature communications}\ }\textbf
  {\bibinfo {volume} {9}},\ \bibinfo {pages} {768} (\bibinfo {year}
  {2018})}\BibitemShut {NoStop}%
\bibitem [{\citenamefont {Frangipane}\ \emph {et~al.}(2018)\citenamefont
  {Frangipane}, \citenamefont {Dell'Arciprete}, \citenamefont {Petracchini},
  \citenamefont {Maggi}, \citenamefont {Saglimbeni}, \citenamefont {Bianchi},
  \citenamefont {Vizsnyiczai}, \citenamefont {Bernardini},\ and\ \citenamefont
  {Di~Leonardo}}]{mona}%
  \BibitemOpen
  \bibfield  {author} {\bibinfo {author} {\bibfnamefont {G.}~\bibnamefont
  {Frangipane}}, \bibinfo {author} {\bibfnamefont {D.}~\bibnamefont
  {Dell'Arciprete}}, \bibinfo {author} {\bibfnamefont {S.}~\bibnamefont
  {Petracchini}}, \bibinfo {author} {\bibfnamefont {C.}~\bibnamefont {Maggi}},
  \bibinfo {author} {\bibfnamefont {F.}~\bibnamefont {Saglimbeni}}, \bibinfo
  {author} {\bibfnamefont {S.}~\bibnamefont {Bianchi}}, \bibinfo {author}
  {\bibfnamefont {G.}~\bibnamefont {Vizsnyiczai}}, \bibinfo {author}
  {\bibfnamefont {M.~L.}\ \bibnamefont {Bernardini}}, \ and\ \bibinfo {author}
  {\bibfnamefont {R.}~\bibnamefont {Di~Leonardo}},\ }\href@noop {} {\bibfield
  {journal} {\bibinfo  {journal} {arXiv preprint arXiv:1802.01156}\ } (\bibinfo
  {year} {2018})}\BibitemShut {NoStop}%
\bibitem [{\citenamefont {Lozano}\ \emph {et~al.}(2016)\citenamefont {Lozano},
  \citenamefont {Ten~Hagen}, \citenamefont {L{\"o}wen},\ and\ \citenamefont
  {Bechinger}}]{bechi}%
  \BibitemOpen
  \bibfield  {author} {\bibinfo {author} {\bibfnamefont {C.}~\bibnamefont
  {Lozano}}, \bibinfo {author} {\bibfnamefont {B.}~\bibnamefont {Ten~Hagen}},
  \bibinfo {author} {\bibfnamefont {H.}~\bibnamefont {L{\"o}wen}}, \ and\
  \bibinfo {author} {\bibfnamefont {C.}~\bibnamefont {Bechinger}},\ }\href@noop
  {} {\bibfield  {journal} {\bibinfo  {journal} {Nature communications}\
  }\textbf {\bibinfo {volume} {7}},\ \bibinfo {pages} {12828} (\bibinfo {year}
  {2016})}\BibitemShut {NoStop}%
\bibitem [{\citenamefont {Stenhammar}\ \emph {et~al.}(2016)\citenamefont
  {Stenhammar}, \citenamefont {Wittkowski}, \citenamefont {Marenduzzo},\ and\
  \citenamefont {Cates}}]{stenhammar2016light}%
  \BibitemOpen
  \bibfield  {author} {\bibinfo {author} {\bibfnamefont {J.}~\bibnamefont
  {Stenhammar}}, \bibinfo {author} {\bibfnamefont {R.}~\bibnamefont
  {Wittkowski}}, \bibinfo {author} {\bibfnamefont {D.}~\bibnamefont
  {Marenduzzo}}, \ and\ \bibinfo {author} {\bibfnamefont {M.~E.}\ \bibnamefont
  {Cates}},\ }\href@noop {} {\bibfield  {journal} {\bibinfo  {journal} {Science
  advances}\ }\textbf {\bibinfo {volume} {2}},\ \bibinfo {pages} {e1501850}
  (\bibinfo {year} {2016})}\BibitemShut {NoStop}%
\bibitem [{\citenamefont {Geiseler}\ \emph {et~al.}(2016)\citenamefont
  {Geiseler}, \citenamefont {H{\"a}nggi}, \citenamefont {Marchesoni},
  \citenamefont {Mulhern},\ and\ \citenamefont {Savel'ev}}]{hanggi}%
  \BibitemOpen
  \bibfield  {author} {\bibinfo {author} {\bibfnamefont {A.}~\bibnamefont
  {Geiseler}}, \bibinfo {author} {\bibfnamefont {P.}~\bibnamefont
  {H{\"a}nggi}}, \bibinfo {author} {\bibfnamefont {F.}~\bibnamefont
  {Marchesoni}}, \bibinfo {author} {\bibfnamefont {C.}~\bibnamefont {Mulhern}},
  \ and\ \bibinfo {author} {\bibfnamefont {S.}~\bibnamefont {Savel'ev}},\
  }\href@noop {} {\bibfield  {journal} {\bibinfo  {journal} {Physical Review
  E}\ }\textbf {\bibinfo {volume} {94}},\ \bibinfo {pages} {012613} (\bibinfo
  {year} {2016})}\BibitemShut {NoStop}%
\bibitem [{\citenamefont {Yellen}\ and\ \citenamefont
  {Virgin}(2009)}]{yellen2009nonlinear}%
  \BibitemOpen
  \bibfield  {author} {\bibinfo {author} {\bibfnamefont {B.~B.}\ \bibnamefont
  {Yellen}}\ and\ \bibinfo {author} {\bibfnamefont {L.~N.}\ \bibnamefont
  {Virgin}},\ }\href@noop {} {\bibfield  {journal} {\bibinfo  {journal}
  {Physical Review E}\ }\textbf {\bibinfo {volume} {80}},\ \bibinfo {pages}
  {011402} (\bibinfo {year} {2009})}\BibitemShut {NoStop}%
\bibitem [{\citenamefont {Tierno}\ and\ \citenamefont
  {Straube}(2016)}]{tierno2016transport}%
  \BibitemOpen
  \bibfield  {author} {\bibinfo {author} {\bibfnamefont {P.}~\bibnamefont
  {Tierno}}\ and\ \bibinfo {author} {\bibfnamefont {A.~V.}\ \bibnamefont
  {Straube}},\ }\href@noop {} {\bibfield  {journal} {\bibinfo  {journal} {The
  European Physical Journal E}\ }\textbf {\bibinfo {volume} {39}},\ \bibinfo
  {pages} {54} (\bibinfo {year} {2016})}\BibitemShut {NoStop}%
\bibitem [{\citenamefont {S{\'a}ndor}\ \emph {et~al.}(2017)\citenamefont
  {S{\'a}ndor}, \citenamefont {Lib{\'a}l}, \citenamefont {Reichhardt},\ and\
  \citenamefont {Reichhardt}}]{sandor2017collective}%
  \BibitemOpen
  \bibfield  {author} {\bibinfo {author} {\bibfnamefont {C.}~\bibnamefont
  {S{\'a}ndor}}, \bibinfo {author} {\bibfnamefont {A.}~\bibnamefont
  {Lib{\'a}l}}, \bibinfo {author} {\bibfnamefont {C.}~\bibnamefont
  {Reichhardt}}, \ and\ \bibinfo {author} {\bibfnamefont {C.~O.}\ \bibnamefont
  {Reichhardt}},\ }\href@noop {} {\bibfield  {journal} {\bibinfo  {journal}
  {Physical Review E}\ }\textbf {\bibinfo {volume} {95}},\ \bibinfo {pages}
  {012607} (\bibinfo {year} {2017})}\BibitemShut {NoStop}%
\bibitem [{\citenamefont {Marconi}\ \emph {et~al.}(2017)\citenamefont
  {Marconi}, \citenamefont {Sarracino}, \citenamefont {Maggi},\ and\
  \citenamefont {Puglisi}}]{marconi2017self}%
  \BibitemOpen
  \bibfield  {author} {\bibinfo {author} {\bibfnamefont {U.~M.~B.}\
  \bibnamefont {Marconi}}, \bibinfo {author} {\bibfnamefont {A.}~\bibnamefont
  {Sarracino}}, \bibinfo {author} {\bibfnamefont {C.}~\bibnamefont {Maggi}}, \
  and\ \bibinfo {author} {\bibfnamefont {A.}~\bibnamefont {Puglisi}},\
  }\href@noop {} {\bibfield  {journal} {\bibinfo  {journal} {Physical Review
  E}\ }\textbf {\bibinfo {volume} {96}},\ \bibinfo {pages} {032601} (\bibinfo
  {year} {2017})}\BibitemShut {NoStop}%
\bibitem [{\citenamefont {Goldstein}(1996)}]{goldstein1996traveling}%
  \BibitemOpen
  \bibfield  {author} {\bibinfo {author} {\bibfnamefont {R.~E.}\ \bibnamefont
  {Goldstein}},\ }\href@noop {} {\bibfield  {journal} {\bibinfo  {journal}
  {Physical review letters}\ }\textbf {\bibinfo {volume} {77}},\ \bibinfo
  {pages} {775} (\bibinfo {year} {1996})}\BibitemShut {NoStop}%
\bibitem [{\citenamefont {Koumakis}\ \emph {et~al.}(2014)\citenamefont
  {Koumakis}, \citenamefont {Maggi},\ and\ \citenamefont
  {Di~Leonardo}}]{barriers}%
  \BibitemOpen
  \bibfield  {author} {\bibinfo {author} {\bibfnamefont {N.}~\bibnamefont
  {Koumakis}}, \bibinfo {author} {\bibfnamefont {C.}~\bibnamefont {Maggi}}, \
  and\ \bibinfo {author} {\bibfnamefont {R.}~\bibnamefont {Di~Leonardo}},\
  }\href@noop {} {\bibfield  {journal} {\bibinfo  {journal} {Soft matter}\
  }\textbf {\bibinfo {volume} {10}},\ \bibinfo {pages} {5695} (\bibinfo {year}
  {2014})}\BibitemShut {NoStop}%
\bibitem [{\citenamefont {Berg}(2008)}]{berg2008coli}%
  \BibitemOpen
  \bibfield  {author} {\bibinfo {author} {\bibfnamefont {H.~C.}\ \bibnamefont
  {Berg}},\ }\href@noop {} {\emph {\bibinfo {title} {E. coli in Motion}}}\
  (\bibinfo  {publisher} {Springer Science \& Business Media},\ \bibinfo {year}
  {2008})\BibitemShut {NoStop}%
\bibitem [{\citenamefont {Romanczuk}\ \emph {et~al.}(2012)\citenamefont
  {Romanczuk}, \citenamefont {B{\"a}r}, \citenamefont {Ebeling}, \citenamefont
  {Lindner},\ and\ \citenamefont {Schimansky-Geier}}]{romanczuk2012active}%
  \BibitemOpen
  \bibfield  {author} {\bibinfo {author} {\bibfnamefont {P.}~\bibnamefont
  {Romanczuk}}, \bibinfo {author} {\bibfnamefont {M.}~\bibnamefont {B{\"a}r}},
  \bibinfo {author} {\bibfnamefont {W.}~\bibnamefont {Ebeling}}, \bibinfo
  {author} {\bibfnamefont {B.}~\bibnamefont {Lindner}}, \ and\ \bibinfo
  {author} {\bibfnamefont {L.}~\bibnamefont {Schimansky-Geier}},\ }\href@noop
  {} {\bibfield  {journal} {\bibinfo  {journal} {The European Physical Journal
  Special Topics}\ }\textbf {\bibinfo {volume} {202}},\ \bibinfo {pages} {1}
  (\bibinfo {year} {2012})}\BibitemShut {NoStop}%
\bibitem [{\citenamefont {Maggi}\ \emph {et~al.}(2015)\citenamefont {Maggi},
  \citenamefont {Marconi}, \citenamefont {Gnan},\ and\ \citenamefont
  {Di~Leonardo}}]{mucna}%
  \BibitemOpen
  \bibfield  {author} {\bibinfo {author} {\bibfnamefont {C.}~\bibnamefont
  {Maggi}}, \bibinfo {author} {\bibfnamefont {U.~M.~B.}\ \bibnamefont
  {Marconi}}, \bibinfo {author} {\bibfnamefont {N.}~\bibnamefont {Gnan}}, \
  and\ \bibinfo {author} {\bibfnamefont {R.}~\bibnamefont {Di~Leonardo}},\
  }\href@noop {} {\bibfield  {journal} {\bibinfo  {journal} {Scientific
  reports}\ }\textbf {\bibinfo {volume} {5}},\ \bibinfo {pages} {10742}
  (\bibinfo {year} {2015})}\BibitemShut {NoStop}%
\bibitem [{\citenamefont {Chatterjee}\ \emph {et~al.}(2014)\citenamefont
  {Chatterjee}, \citenamefont {Chatterjee}, \citenamefont {Pradhan},\ and\
  \citenamefont {Manna}}]{chatterjee2014interacting}%
  \BibitemOpen
  \bibfield  {author} {\bibinfo {author} {\bibfnamefont {R.}~\bibnamefont
  {Chatterjee}}, \bibinfo {author} {\bibfnamefont {S.}~\bibnamefont
  {Chatterjee}}, \bibinfo {author} {\bibfnamefont {P.}~\bibnamefont {Pradhan}},
  \ and\ \bibinfo {author} {\bibfnamefont {S.}~\bibnamefont {Manna}},\
  }\href@noop {} {\bibfield  {journal} {\bibinfo  {journal} {Physical Review
  E}\ }\textbf {\bibinfo {volume} {89}},\ \bibinfo {pages} {022138} (\bibinfo
  {year} {2014})}\BibitemShut {NoStop}%
\end{thebibliography}%

\end{document}